\documentclass[aps,pra,reprint,superscriptaddress]{revtex4-2}

\usepackage{xr}
\usepackage{graphicx}
\usepackage{dcolumn}
\usepackage{bm}
\usepackage[mathlines]{lineno}
\usepackage{titlesec}
\usepackage{url}

\titlespacing*{\section} {0pt}{5ex}{2ex}
\usepackage{hyphenat}
\makeatletter

\begin{document}


\title{Efficient and Tunable Photochemical Charge Transfer via Long-Lived Bloch Surface Wave Polaritons \footnote{Error!}}

\author{Kamyar Rashidi}
\altaffiliation{Photonics Initiative, Advanced Science Research Center, City University of New York, New York, New York 10031, USA}

\altaffiliation{Department of Physics, Graduate Center, City University of New York, New York, New York 10016, USA}
 
\author{Evripidis Michail}%
\affiliation{Photonics Initiative, Advanced Science Research Center, City University of New York, New York, New York 10031, USA}
 
\affiliation{Department of Physics, Graduate Center, City University of New York, New York, New York 10016, USA}

\author{Bernardo Salcido-Santacruz}%
\affiliation{Photonics Initiative, Advanced Science Research Center, City University of New York, New York, New York 10031, USA}

\affiliation{Department of Chemistry, Graduate Center, City University of New York, New York, New York 10016, USA}

\author{Yamuna Paudel}%
\affiliation{Photonics Initiative, Advanced Science Research Center, City University of New York, New York, New York 10031, USA}

\affiliation{Department of Physics, Graduate Center, City University of New York, New York, New York 10016, USA}
 
\author{Vinod M. Menon}%
\affiliation{Department of Physics, Graduate Center, City University of New York, New York, New York 10016, USA}

\affiliation{Department of Physics, City College of New York, New York, New York 10031, USA}%

\author{Matthew Y. Sfeir} \email{msfeir@gc.cuny.edu}
\affiliation{Photonics Initiative, Advanced Science Research Center, City University of New York, New York, New York 10031, USA}
 
 \affiliation{Department of Physics, Graduate Center, City University of New York, New York, New York 10016, USA}

\date{\today}

\begin{abstract}
Achieving precise control of photoinduced molecular charge transfer reactions underpins key emerging technologies. As such, the use of hybrid light-matter molecular exciton-polariton states has been proposed as a scheme to directly modify the efficiency and rate of such reactions. However, the efficacy of polariton-driven photochemistry remains an open question. Here, we demonstrate conditions under which photoinduced polaritonic charge transfer can be achieved and directly visualized using momentum resolved ultrafast spectroscopy. Key conditions for charge transfer are satisfied using Bloch surface wave polaritons, which exhibit favorable dispersion characteristics that permit the selective pumping of hybrid states with long lifetimes (100-400 fs) that permit vibrationally assisted molecular charge transfer. Using this approach, we tune the energetic driving force for charge separation, reducing it by as much as 0.5 eV compared to the bare exciton. These results establish that tunable and efficient polariton-driven molecular charge transfer is indeed possible using carefully considered photonic systems. 

\end{abstract}

\maketitle

Organic exciton polaritons emerge due to strong coupling between the excitonic states of organic materials and the electromagnetic modes of optical cavities, creating hybrid quasiparticles with both matter and photon characteristics\cite{weisbuch1992observation, hopfield1958theory}. A key advantage of organic materials in strong light-matter coupling applications is that they enable the investigation of novel quantum states at room temperature\cite{ghosh2022microcavity}, including such phenomena as polariton condensation\cite{kasprzak2006bose}, superfluidity\cite{amo2009superfluidity}, and soliton propagation\cite{lagoudakis2008quantized}. In addition, due to their strong absorption, large delocalization, and tunable energetics, organic exciton-polaritons have also been widely proposed as attractive candidates for applications involving photoinduced charge transfer (CT), such as optoelectronics, spin conversion processes, and isomerization reactions\cite{zasedatelev2019room,hutchison2013tuning,orgiu2015conductivity, rozenman2018long, coles2014polariton, du2018theory, georgiou2021ultralong,liu2022photocurrent, nikolis2019strong, bhuyan2023rise}. 

\begin{figure*}
    \centering
\includegraphics[width=1\linewidth]{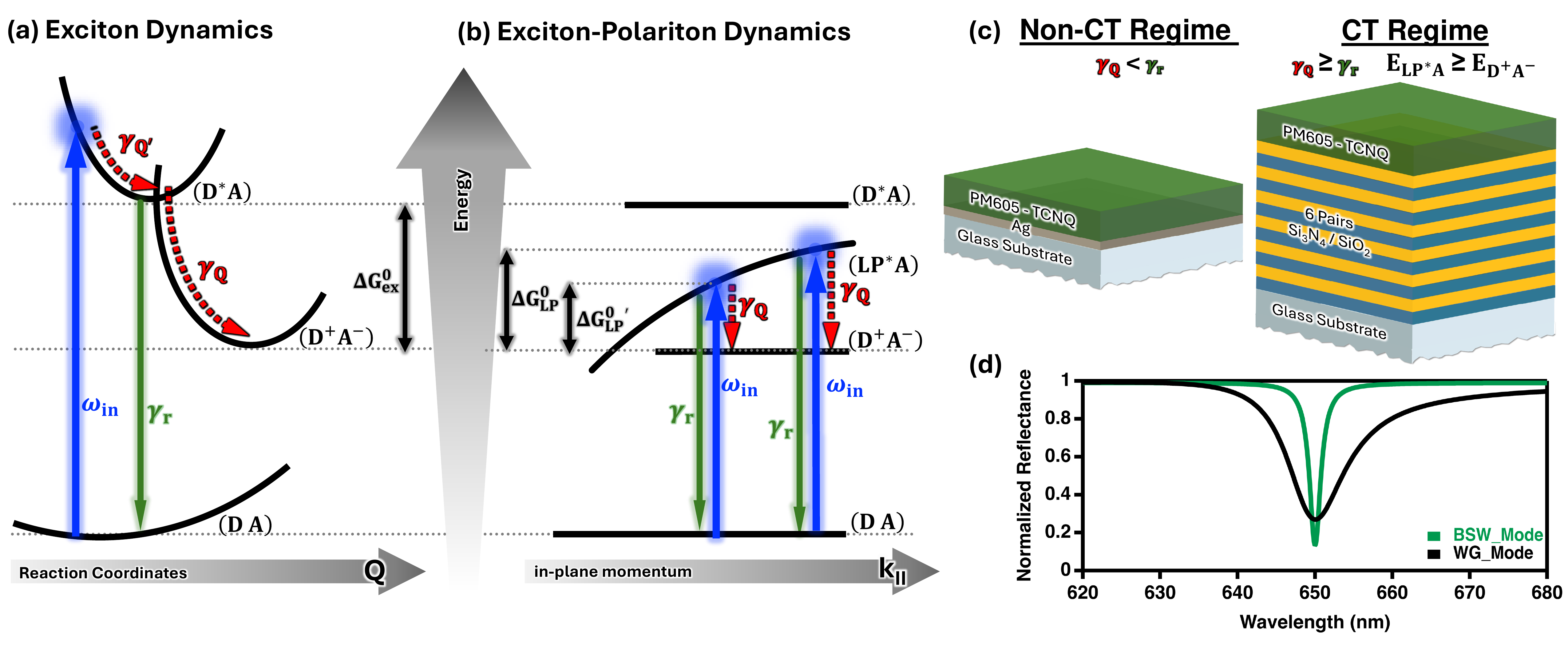}    
\caption{\textbf {Charge Transfer (CT) Processes in Donor-Acceptor Systems: Bare vs. Metal and DBR Cavities.} Schematic representation of photon absorption from the ground state (GS, DA) with energy $\hbar\omega_{\text{in}}$ to the excited state of the (a) bare donor-acceptor system (D*A) and (b) cavity donor-acceptor system (LP*A). Two competing pathways of relaxation to the ground state and transfer to the charge-transfer state (D$^+$A$^-$) are shown with their respective rate constants ($\gamma_{\text{r}}$ and $\gamma_{\text{Q}}$) where the arrows represent population transfer. For bare excitons, the driving force (\(\Delta G^{0}_{\text{ex}}\)) remains constant and relatively high. However, with polaritons, this value can be adjusted to various lower levels, represented by \(\Delta G^{0}_{\text{LP}}\) and \(\Delta G^{0}_{\text{LP}'}\).
(c) Left: A metal cavity structure composed of a 150 nm PM605-TCNQ film spin-coated onto a 30 nm silver layer. Due to the lower lifetime (lower Q-factor) of the polariton state, electron relaxation to the ground state occurs before charge transfer can happen. (c) Right: A photonic structure composed of an 80 nm PM605-TCNQ film spin-coated onto six pairs of Si$_3$N$_4$ (110 nm)/SiO$_2$ (146 nm) dielectric Bragg reflectors. The longer lifetime (higher Q-factor) of the polariton state in this structure allows for charge transfer to occur.(d) Reflection spectra of the uncoupled waveguide (WG) mode in metallic cavity and BSW in distributed Bragg reflector (DBR) cavities were obtained using the transfer matrix method. The Q-factors for the WG mode and BSW were measured as 73 and 366, respectively.}
  \label{fgr:figri1}
\end{figure*}

The potential advantages of polariton photochemistry have been widely championed \cite{mandal2019investigating,yuen2019polariton,hutchison2012modifying,munkhbat2018suppression}. In the field of molecular photochemistry, a great deal of effort is devoted to engineering the reactivity of the excited state by tuning its potential energy surface and the character of the its wavefunction. For example, the conversion from a bound excitonic to an unbound charge separated state requires a energetic driving force\cite{ward2015impact,unger2017impact} (Fig.~\ref{fgr:figri1}a) that is generally provided in the form of a chemical potential difference between an excited state of an absorbing donor molecule and an unoccupied molecular orbital of an acceptor compound. In this context, polariton-driven photochemistry would uniquely enable a means of dynamically tuning excited state potentials via dispersion of the hybrid state without modification of the molecular structure (Fig.~\ref{fgr:figri1}b). This potentially allows for the determination of electronic and kinetic factors that dictate charge transfer reactions\cite{paul2010evidence}, photoisomerizations\cite{delesma2021photoisomerization}, and spin conversion processes among others\cite{schafer2019modification,kim2024understanding}. If realized, the phenomena of tunable polariton photochemistry could have an impact analogous to the scientific revolution enabled by the size and shape dependent properties of quantum-confined nanomaterials.  

However, fundamental challenges have so far precluded the realization of dynamically tunable exciton polariton-based photochemical systems. The lack of a consensus as to the efficacy of organic exciton-polariton photochemistry is rooted in several factors that affect the understanding of their excited state dynamics and has hindered the reproducibility of many studies. One issue is that in molecular systems, charge-transfer processes are generally understood within the context of Marcus Theory, which described how changes in the spatial charge distribution are stabilized by corresponding changes in the nuclear coordinates (Fig.~\ref{fgr:figri1}a)\cite{deibel2010role,marcus1956theory}. As such, the rate of CT is fundamentally limited by the atomic motion rather than the electron tunneling time\cite{ETbook}. In other words, for intermolecular CT to proceed efficiently, the lifetime of the excited donor state ($1/\gamma_{r}$) must exceed the timescale for nuclear reorganization ($1/\gamma_{Q}$), whose lower limit is on the order of 100 fs \cite{hou_incoherent_2023}. Notably, this condition is not always met under strong coupling conditions. For example, typical all-metal microcavities exhibit fast momentum scattering ($\gamma_{k}$) and photon lifetimes of $< 25$ fs (Fig.~\ref{fgr:figri1}c). These short lifetimes likely preclude efficient CT, even under conditions where selective exciton-polariton excitation is achieved \cite{baranov2018novel,michail2024addressing}.

Additionally, to understand the dynamics of polariton photochemistry, it is necessary to distinguish between CT resulting from hybrid states compared to matter states\cite{climent2022not,grafton2021excited}. This is particularly challenging in organic systems where a large amount of inhomogeneous broadening together with strong electron-vibrational coupling imparts significant oscillator strength to reservoir states\cite{khazanov2023embrace,dutta_thermal_2024}. These states are formally linear combinations of exciton basis states, and it has been experimentally shown that their energies, potential energy surfaces, and dynamics are largely identical to bare, uncoupled excitons\cite{khazanov2023embrace, liu2021ultrafast, dutta_thermal_2024}. However, distinguishing between polariton and reservoir state dynamics can be very challenging since energy storage in reservoir states leads to spurious transient signals at polariton resonances. These signals result from small changes in the photonic properties of the system, but are frequently misinterpreted and incorrectly used to assign polariton lifetimes\cite{litinskaya2004fast,schwartz2013polariton, lodden2010electrical}. This problem can be addressed by selecting a photonic system with a dispersion suitable for selective excitation of polaritons. For example, we have shown in recent work that the steep dispersion associated with surface and waveguide modes permit selective excitation of hybrid states via their energy- and momentum-resonance conditions and permits a direct measurement of its decay dynamics\cite{michail2024addressing}. Still, it remains a grand challenge to demonstrate that the rate and efficiency of photochemical processes based on CT are inherently impacted by the presence of hybrid polariton states.

We posit that to fully identify and understand the dynamics of polariton-driven photochemical processes, it is therefore necessary to achieve both selective excitation of exciton-polaritons and exciton-polariton lifetimes that exceed the characteristic timescale for nuclear relaxation. To test this, we leverage the favorable properties of organic materials strongly coupled to Bloch surface waves (BSW). The advantages of this system has been recently demonstrated to include their exceptionally high group velocity, prolonged lifetime, and minimal dissipation that have led to long propagation distances at room temperature\cite{liscidini2011guided, pirotta2014strong, barachati2018interacting, bhaskar2020bloch, mousavi2022computational, lerario2017high}. For the purposes of CT, the simple geometry and long lifetimes are promising for mechanistic studies of polariton mediated charge transfer. Furthermore, we are able to take advantage of the dispersion of BSW-polaritons to selectively excite the lower polariton (LP) state under energy- and momentum-matching conditions. Here we show that polariton-enhanced CT can indeed be achieved under these specific conditions using a strongly coupled organic donor molecular and a molecular acceptor through the identification of ion states that are resonantly generated on ultrafast timescales. Importantly, CT is not achieved in analogous metallic cavities with similar dispersion properties due to their short lifetimes relative to nuclear relaxation. We show that the dynamics of BSW-polaritons can be observed through both LP and Internal Optical Modes (IOMs) and that significant quenching of the polariton emission signal is observed in the CT regime. Ultimately, we achieve efficient and tunable CT using the LP-BSW Mode and show that the overpotential for CT can be reduced by as much as 0.5 eV.

\section{Photonic structure design considerations}
Our photonic structure with long-lived exciton-polariton states consists of an 80 nm layer of organic dye encapsulated in a polymethyl methacrylate (PMMA) matrix that is deposited on the surface of a one dimensional photonic crystal comprised of six pairs of alternating  $\text{Si}_3\text{N}_4$ (110 nm) / $\text{SiO}_2$ (146 nm) layers (Fig.~\ref{fgr:figri1}c (right)). The organic layer consists of either Pyrromethene 605 donor molecules (PM605), Tetracyanoquinodimethane acceptor molecules (TCNQ), or a mixture of both the donor and acceptor in a 1:1 ratio (PM605-TCNQ 1:1). A BSW mode located within the photonic bandgap gives rise to a confined electric field in the immediate vicinity of the surface. Spatial overlap of the confined electromagnetic field and the strongly absorbing organic dye leads to strong light-matter coupling and the formation of exciton-polaritons with a large Rabi splitting ($\hbar\Omega \propto (N/V_m)^{1/2}$), where $V_m$ is the effective mode volume and $N$ is the number of coupled molecules. In the strong coupling regime, the polariton lifetime will be limited by the uncoupled photonic lifetime through \(\tau_{\text{xp}}^{-1} = A \tau_{\text{p}}^{-1} + B \tau_{\text{x}}^{-1}\), where \(\tau_{\text{p}}\) and \(\tau_{\text{x}}\) are the lifetimes of the uncoupled photonic and excitonic states, and \(A\) and \(B\) are their respective fractions in the wavefunction. Here, the quality factor (\(Q\)-factor) of the uncoupled photonic BSW mode, in which the organic layer consists of only the PMMA matrix, is calculated using transfer matrix methods to be approximately 366, corresponding to $\tau_{p} > 125$ fs (Fig.~\ref{fgr:figri1}d). This lifetime is significantly longer than corresponding modes in other common photonic systems, such as surface plasmon polaritons where $\tau_{p} < 10$ fs as well as metal microcavity and waveguide modes in which $\tau_{p} < 25$ fs\cite{michail2024addressing}. These differences in the relaxation times are extremely significant in the context of molecular CT. For photonic systems in which the lifetime of the photonic state is shorter than what is required for vibrationally-mediated charge transfer, charge separation is suppressed due to rapid ground state recovery (Fig.~\ref{fgr:figri1}c (left)). In contrast, BSW and other high-Q photonic systems exhibit higher $\tau_{p}$, such that incoherent CT processes become accessible (Fig.~\ref{fgr:figri1}c (right)).  As such, similar to the case of a bare exciton, a long LP lifetime is necessary to facilitate CT from a hybrid light-matter state.  
\section{Charge transfer in the bare D-A film}

 \begin{figure}
    \centering    \includegraphics[width=0.9\linewidth]{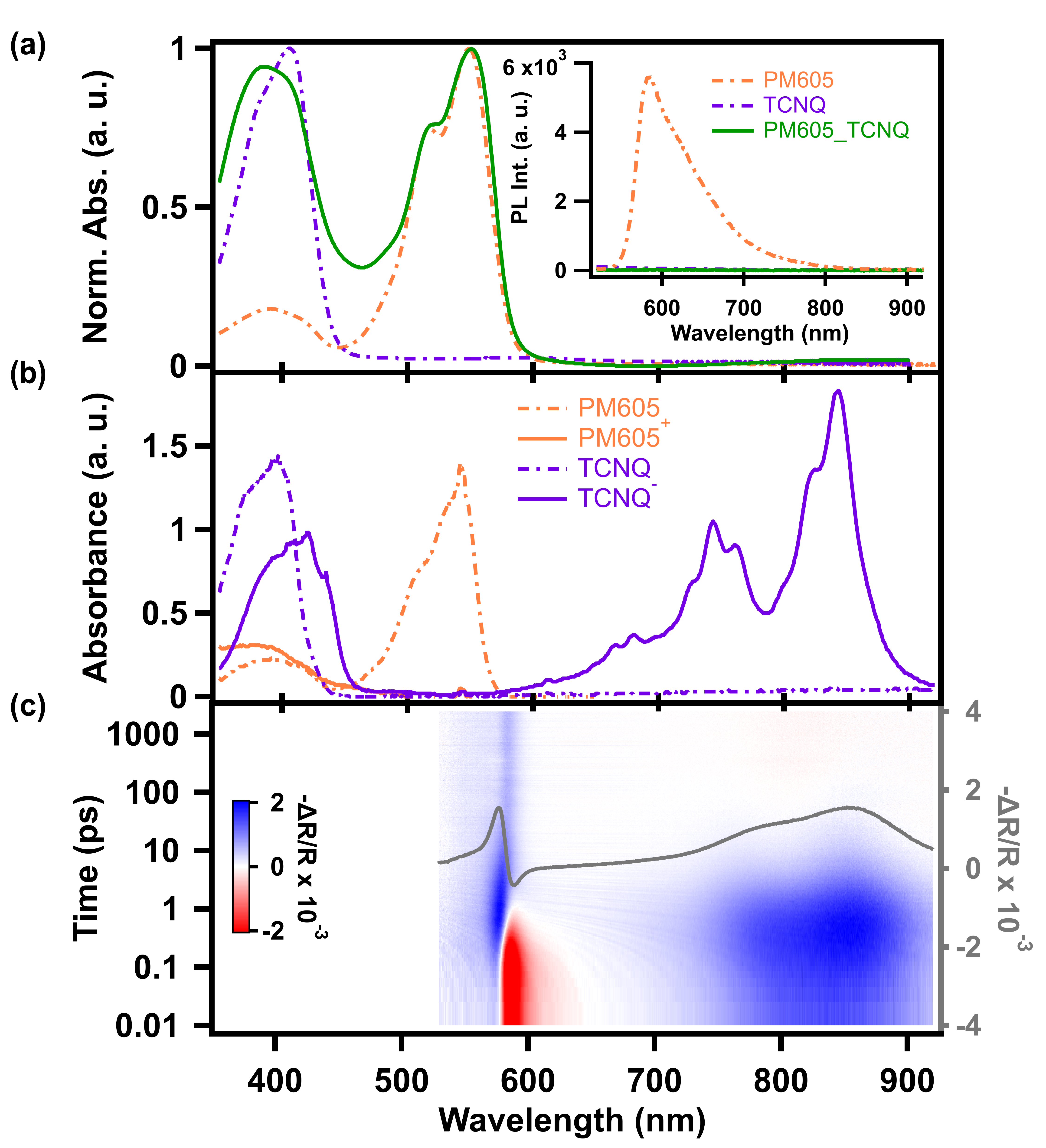}    
  \caption{\textbf{Highlighting CT Signatures in PM605-TCNQ Blend Film.} (a) Absorption of PM605, TCNQ, and their 1:1 mixture PM605-TCNQ. The inset displays the emission spectra, where emission is quenched for the PM605-TCNQ blend film. (b) Steady-state absorption spectra of PM605, PM605\(^+\), TCNQ, and TCNQ\(^-\) in acetonitrile measured using spectroelectrochemical techniques. TCNQ\(^-\) exhibits two additional anion absorption peaks at 744 nm and 843 nm. PM605 has a peak at 542 nm, which is quenched for its cation (PM605\(^+\)). (c) Transient reflection spectra of the PM605-TCNQ 1:1 mixture. The secondary vertical axis represents the average transient reflection from 0.1 to 4 ps, revealing excited state absorption around the anion absorption, as well as quenching of the ground state bleach near the PM605 absorption.}
  \label{fgr:abss}
\end{figure}

Steady-state and transient optical spectroscopy combined with spectroelectrochemical measurements confirm that rapid and efficient formation of a charge separated state occurs in the PM605-TCNQ blend outside of the cavity. The absorption spectrum of the blend largely resembles a linear combination of the individual components (Fig.~\ref{fgr:abss}a), with the neutral PM605 and TCNQ components identified by their distinct absorption profiles at 550 nm and 400 nm, respectively. Notably, the emission spectra of the PM605-TCNQ blend (inset of Fig.~\ref{fgr:abss}a) show strong quenching relative to a neat PM605 film. The energies of the frontier molecular orbitals determined from cyclic voltammetry suggest that a large ($\sim 700$ meV) energetic driving force facilitates excited state electron transfer from the PM605 exciton to the TCNQ (Fig. S1). The corresponding ion spectra (Fig.~\ref{fgr:abss}b) indicate that TCNQ$^{-}$ exhibits strong characteristic absorption peaks at 744 nm and 843 nm. While PM605$^{+}$ does not exhibit a characteristic transition in our spectral window, the neutral absorption peak of PM605 at 542 nm is largely suppressed upon ionization. 
Transient optical studies performed on films of the individual components and their blend confirm that ultrafast electron transfer dynamics occur in the blend upon photoexcitation (Fig. S2),Fig.~\ref{fgr:abss}c). These measurements (Fig.~\ref{fgr:abss}c) reveal a concurrent decay of ground state bleach of PM605 around 590 nm  and the rise of excited state absorption (ESA) around 780 nm  and 850 nm corresponding to anion formation (TCNQ\(^-\)). From an analysis of the kinetics, we determine that CT occurs within 180 fs resulting in an anion lifetime of 5.41 ps.
\section{Momentum-resolved exciton-polariton dynamics}
\begin{figure*}
    \centering    \includegraphics[width=1\linewidth]{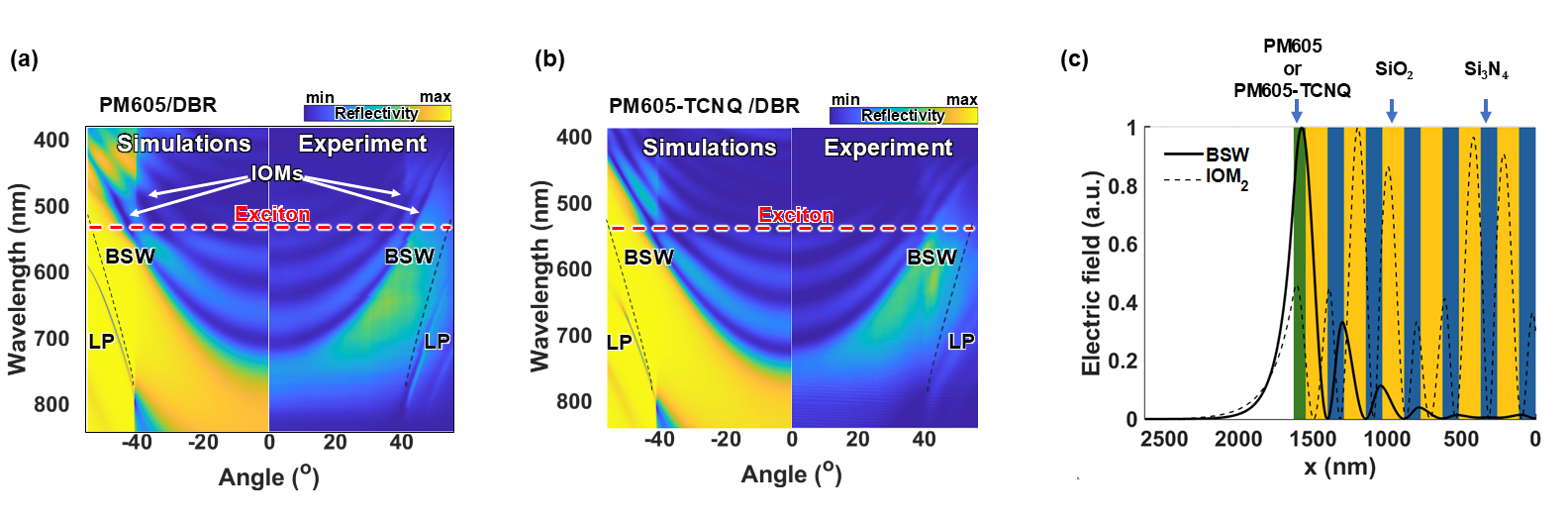}    
\caption{\textbf{Strong coupling between the BSW mode and excitonic states.} Simulated (left) and measured (right) TE-polarized angle-resolved reflection spectra of (a) PM605/DBR and (b) PM605-TCNQ/DBR. The uncoupled exciton and BSW dispersions are represented by dashed red and black lines, respectively. (c) DBR configuration supporting in-plane propagating BSW and IOM$_2$ modes.}

  \label{fgr:figri2}
\end{figure*}

We identify that strong light-matter coupling is achieved between the PM605 exciton and the BSW mode when both the neat PM605 film and the PM605-TCNQ blend are deposited on our DBR structure. The angle-resolved reflectivity of PM605/DBR and PM605-TCNQ/DBR structures (Fig.~\ref{fgr:figri2}a,b) illustrates the avoided crossing between the BSW mode (dashed black lines) and the exciton energy level (dashed red line near 542 nm) of PM605 above the critical angle ($\sim 40 ^\circ$). The resulting LP state (solid black line) can be seen as a dip in the reflectivity in both simulations and experiments. We note that the upper polariton is not readily identifiable within our measurement range. In addition to the BSW surface mode, the DBR exhibits several IOMs above and below the light line, for which the electric field is extended into either air or the dielectric stack (Fig.~\ref{fgr:figri2}c). As we will show later, these modes are weakly coupled to the organic layer and exhibit a modified dispersion due to changes to the effective refractive index of the organic layer in the excited state. The separation between the BSW and the LP indicates the strong coupling regime, satisfying the condition $ \hbar\Omega_R \geq (\Gamma_M + \Gamma_C)/2 $, where $ \hbar\Omega_R $ denotes the Rabi splitting energy (585 meV), while $ \Gamma_M $ and $ \Gamma_C $ represent the full width at half maximum linewidths of the bare molecule exciton (255 meV) and the cavity mode (5.4 meV), respectively. 

\begin{figure*}
    \centering    \includegraphics[width=1\linewidth]{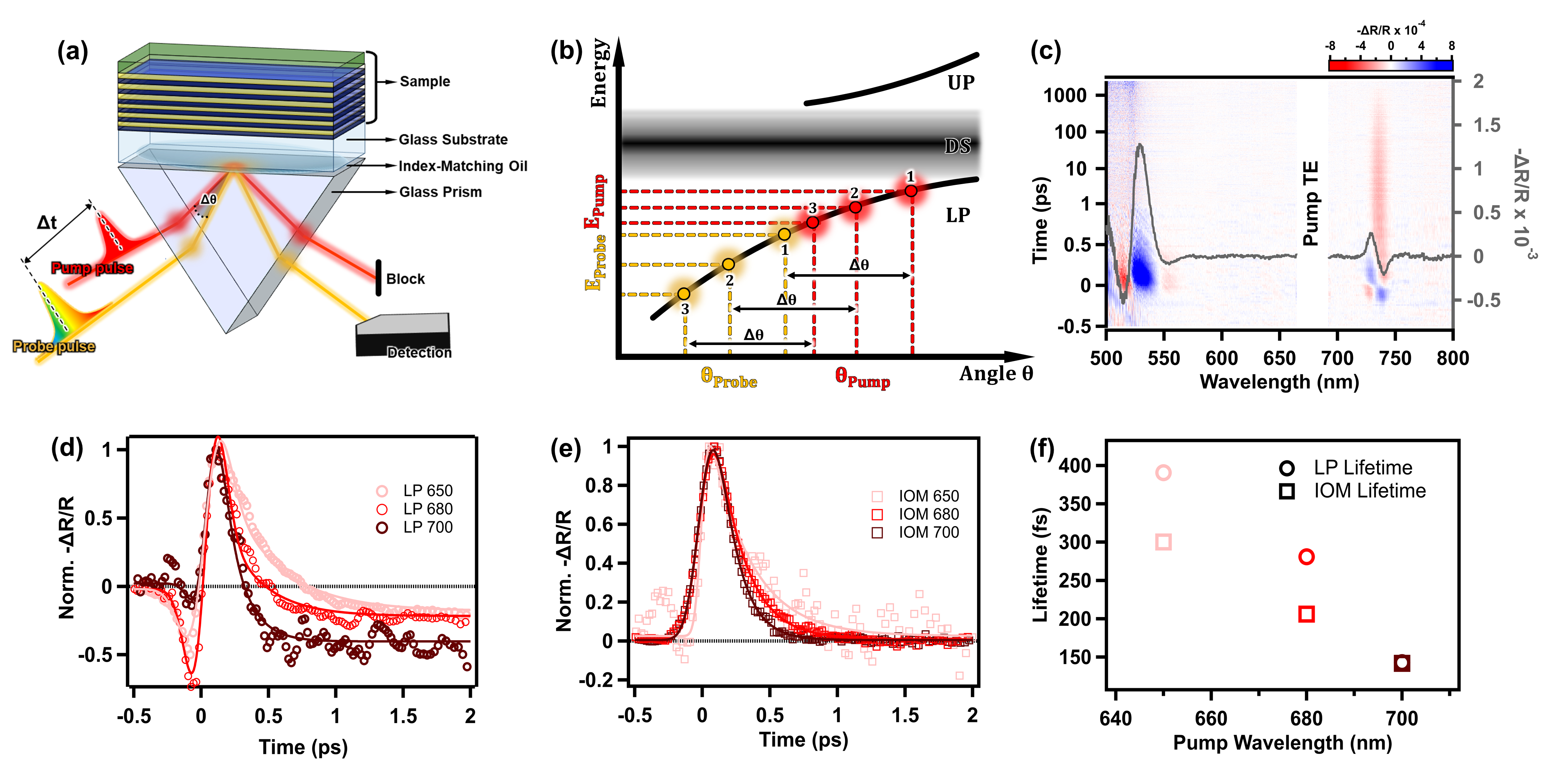}  
\caption{\textbf{Lifetime of BSW polaritons across various points on the LP branch.} (a) Schematic representation of the pump-probe experimental configuration in the Kretschmann-Raether setup. The pump and probe beams are directed onto the sample with an angular difference $\Delta\theta = \theta_{\text{pump}} - \theta_{\text{probe}}$. The probe beam (white light) is delayed by a time $\Delta t$ after the pump. The sample is positioned on top of the prism with index-matching oil, substrate side down. (b) Schematic diagram illustrating the excitation of LPs with three different pump wavelengths (650 nm, 680 nm, and 700 nm) under phase-matching condition (PMC), while maintaining a constant angle between the pump and probe beams ($\Delta\theta = 6.2^\circ$). (c) Transient reflection upon excitation with a 680 nm pump. The secondary vertical axis displays transient spectra within the first 200 fs. (d) Kinetic traces of LP and (e) IOMs for the three different pump wavelengths at the positions of gray dashed lines in (c). (f) Lifetimes of polariton and IOM corresponding to three different excitation wavelengths.}

  \label{fgr:figri4}
\end{figure*}

Selective excitation of the LP in transient optical measurements is accomplished using the Kretschmann-Raether attenuated total reflection approach that enables the excitation of BSWs from the far-field (Fig.~\ref{fgr:figri4}a). Briefly, the optical pump and probe pulses propagate at high angles through a prism coupled to our photonic structure using index-matching oil. The two pulses are spatially overlapped on the active area of our structure with an angular offset of $\Delta\theta = \theta_{\text{pump}} - \theta_{\text{probe}}$ and a delay time of $\Delta t$. The broadband reflected probe signal detects transient changes in the refractive indices of the organic layer. As we have previously shown, selective excitation of LP states occurs under PMCs in which the pump wavelength and angle are tuned to match the dispersion of the hybrid LP mode (Fig. S3)\cite{michail2024addressing}. Direct measurements of the polariton lifetime are enabled by excitation at specific in-plane momenta that correspond to LPs with both significant excitonic character and a large energetic separation from reservoir states. On the other hand, excitation of reservoir states is accomplished by tuning the wavelength of the pump pulse to be resonant with the bare exciton energy. In contrast to polariton states, the excitation of reservoir states occurs for any value of the in-plane momentum of the pump pulse. We conducted our measurements within the low excitation regime (less than 4 $\mu$J/cm\textsuperscript{2}), employing a high-sensitivity transient optical spectrometer recently developed by our group\cite{hall2023optimizing}.

We established that the lifetime of the BSW exciton-polariton in the neat PM605 film exhibits long lifetimes that exceed the CT time constant that we measured in the uncoupled blend film (180 fs). As the lifetime of the exciton-polariton depends on the relative excitonic and photonic fractions, we tuned the energy and in-plane momentum of our pump pulse to resonantly excite various points on the LP dispersion curve. We performed this experiment under PMCs across three distinct wavelengths ($\lambda_{\text{pump}} = 650$ nm, 680 nm, and 700 nm) with a consistent fluence (1 $\mu$J/cm\textsuperscript{2}) and a constant $\Delta\theta = $ 6.2$^{\circ}$ (Fig.~\ref{fgr:figri4}b). After photoexcitation, the transient spectra exhibit a dispersive shape centered at the LP transition energy, indicating a blue-shift of the hybrid mode in the excited state. Representative data with $\lambda_{pump} = $ 680 nm are shown in Fig.~\ref{fgr:figri4}, for which a strong dispersive feature associated with the LP appears near 735 nm. This response is characteristic of strongly-coupled systems and reflects a transient decrease in Rabi splitting due to the photoexcitation of a subset of coupled molecules. This effect is verified with power-dependent measurements, which show that the magnitude of the blue shift increases with increasing pump power (Fig. S4). The lifetime of the LP state is determined by measuring the timescale for the decay of the dispersive signal near the LP (Fig.~\ref{fgr:figri4}d  and f), which yields a decay constant of approximately 390 fs at the 650 nm pump wavelength. Importantly, this value is twice as long as the CT rate uncoupled PM605-TCNQ blend. As the in-plane momentum pump and probe momenta are decreased (shift of LP to longer wavelengths), the lifetime of LP decreases to $\sim 275$ fs at $\lambda_{pump} = $ 680 nm and $\sim 150$ fs at $\lambda_{pump} = $ 700 nm, consistent with greater photonic character. 

There are two other features that are readily apparent in the transient optical data, but are not relevant to the problem of photoinduced CT. At very early times ($<$ 200 fs) the sign of the dispersive signal is reversed, consistent with an optical Stark effect\cite{hayat2012dynamic, lamountain2021valley}. This sign reversal induces an obvious oscillating component in the early time kinetics near the LP resonance (Fig.~\ref{fgr:figri4}d) but doesn't affect our determination of the LP lifetime. In addition, a long-lived, non-dispersive signal with negative $-\Delta R/R$ is observed near the LP resonance that exhibits a lifetime on the order of 200 ps, similar to the bare PM605 exciton (Fig. S2). Notably, the non-dispersive signals scale linearly with pump fluence and are spectrally distinct from the dispersive modes observed for reservoir pumping. Furthermore, they are not observed for an identical film coupled to surface photonic modes of a metal film. As such, we assign this to weakly coupled stimulated emission originating from uncoupled molecules populated by long-range energy transfer. This emissive species likely originates from dye aggregates with red-shifted emission that are observable only due to the low losses and large group velocities of the BSW mode. In this case, the polariton mode acts only as a spectral filter that facilitates the outcoupling of light from the film\cite{simpkins2023comment}. This effect can be clearly seen through a comparison of the steady-state emission to the long-lived component of the transient optical signal, which exhibit nearly identical lineshapes and relative intensities over a series of angles (Fig. S6). As discussed below, these species are also not observed in blends and as such are irrelevant to our discussion of CT dynamics. 

Additional features corresponding to IOMs are observed in the momentum-resolved transient spectra and support our determination of the polariton lifetime. For example, under photoexcitation at 680 nm, we observe an additional dispersion feature near the IOM resonance at 525 nm in addition to the LP feature at 735 nm (Fig.~\ref{fgr:figri4}c). The IOM feature is also short-lived (Fig.~\ref{fgr:figri4}e) and exhibits decay dynamics that nearly identical to the values obtained at the LP probe wavelength (Fig.~\ref{fgr:figri4}f). This response can be readily understood by inspection of the field profile for the IOM at 525 nm (Fig.~\ref{fgr:figri2}c), which has significant spatial overlap with the organic layer. Transient photobleaching of the dye in the excited state modifies its effective refractive index, leading to small blue shifts in the dispersion of the IOM, consistent with previous findings\cite{renken2021untargeted}. As a result, the dynamics of the IOM-associated feature acts as a proxy for the excited state lifetime of the organic layer. As the IOM signal is free from artifacts stemming from optical Stark effects and background emission of weakly coupled species, it permits facile determination of the polariton lifetime. 

\begin{figure}
    \centering    \includegraphics[width=0.98\linewidth]{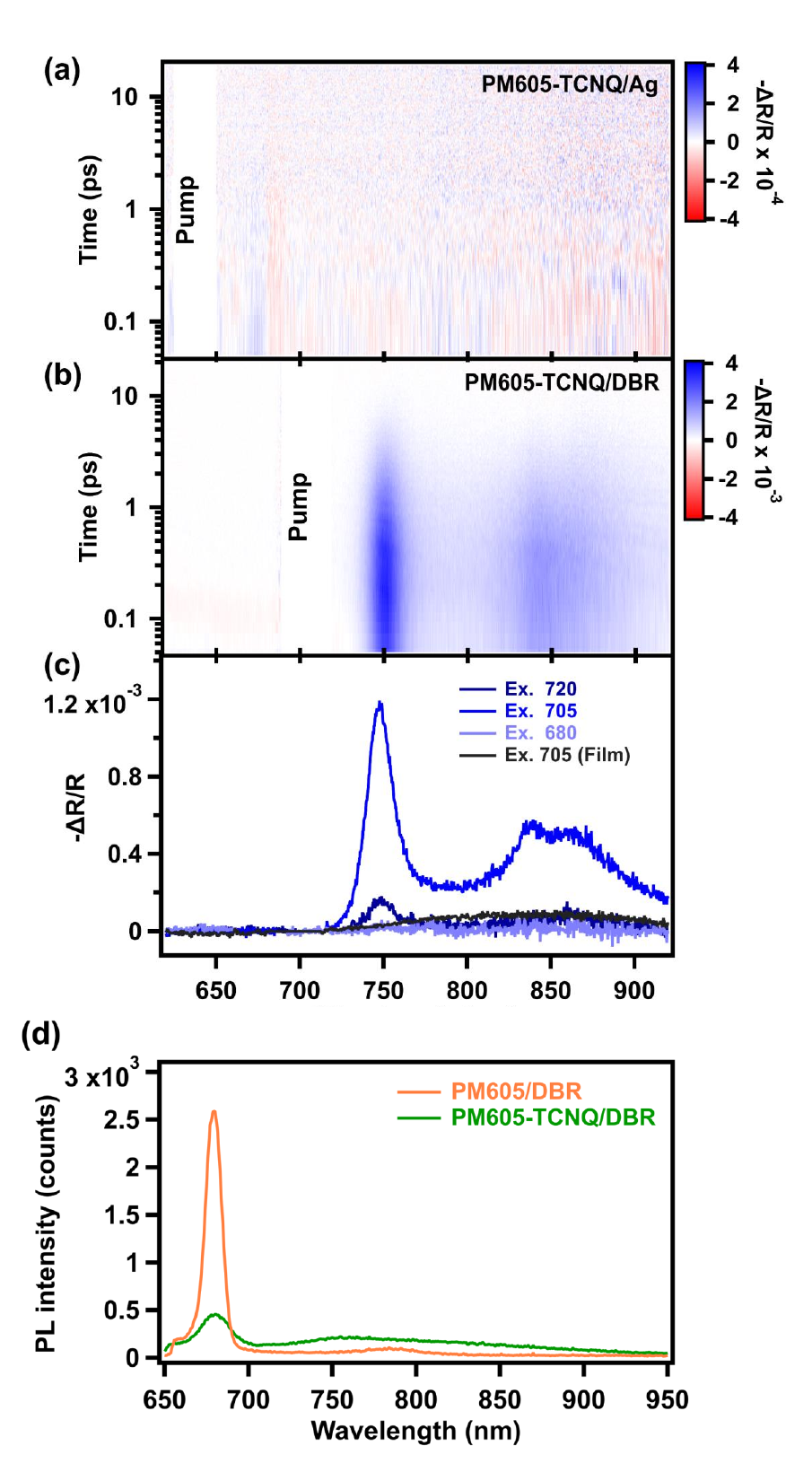}  
    \caption{\textbf{Driving selective and efficient CT with long-lived BSW polaritons}.
(a) Transient reflection data of the strongly coupled PM605-TCNQ/Ag structure under selective LP excitation with suppressed CT. (b) Transient reflection data of PM605-TCNQ/DBR under direct LP excitation, achieving strong CT. (c) Comparative analysis of transient reflection data for PM605-TCNQ film and PM605-TCNQ/DBR under PMC (705 nm) and non-PMCs. (d) PL map showing LP emission quenching for the PM605-TCNQ/DBR structure compared to PM605/DBR under PMC at 633 nm.
}
  \label{fgr:MainFigCT2}
\end{figure}
\section{Tunable charge transfer from exciton-polaritons}

We observe remarkably efficient CT from exciton-polaritons in the PM605-TCNQ/DBR configuration (Fig.~\ref{fgr:MainFigCT2}b) using selective and direct LP excitation, significantly outperforming what is observed with  either an uncoupled blend film or a PM605-TCNQ/Ag photonic structure. For example, there is no indication of the TCNQ\(^-\) signal in the transient optical data when selectively exciting the LP-WG mode at 640 nm in the PM605-TCNQ/Ag structure, consistent with its low quality factor and short lifetime that promotes ground state recombination over CT (Fig.~\ref{fgr:MainFigCT2}a). In the kinetic trace extracted near the LP resonance (Fig. S9), we observe only a small peak with a width that is instrument response function limited, which we assign to the polariton.  In contrast, under the same LP excitation conditions, the BSW hybrid mode on the DBR structure exhibits an additional long-lived feature with a lifetime corresponding to the measured decay of the TCNQ\(^-\) anion (Fig. S9). This effect persists even when the exciting momentum states have larger energetic separation from the reservoir states. Under PMCs at 705 nm, strong anion peaks at 750 nm and 850 nm can be seen in the transient optical data, which can be readily assigned to TCNQ\(^-\) states formed from photoexcited CT. We note that the spectral shape of the anion peaks is modified by the reflectivity spectrum of the DBR in the NIR, such that the strongest features are aligned with the modes of the photonic system. 

Importantly, the intensity of the anion peaks is the highest when  the pump matches the energy- and momentum-resonance conditions of the LP. For pumping with constant momentum and a detuned energy from the polariton dispersion ((Fig.~\ref{fgr:MainFigCT2}d), PMCs are not satisfied for LP excitation and the intensity of the TCNQ\(^-\) anion decreases significantly. We note that the response of the neat blend film, uncoupled to the cavity, is minimal at these pumping wavelengths indicating that hybrid mode plays a role in facilitating CT. Further evidence for the direct involvement of the hybrid mode in the CT process is obtained from photoluminescence measurements (Fig.~\ref{fgr:MainFigCT2}d) in which a marked decrease of the LP emission is observed in the PM605-TCNQ/DBR structure compared to PM605/DBR under PMCs. This effect suggests quenching of the LP emission due to a competing CT channel. The magnitude of the PL quenching ($\sim 20 \%$) permits a rough estimate of the CT time of $\sim$ 85 fs using the lifetime of the LP in the neat PM605 film under identical pumping conditions ($\sim 350$ fs). Consistent with this timescale, our data exhibit an instrument response limited (120 fs) rise in the anion feature. Similarly, it also suggests an explanation for the lack of an observed CT signal in the PM605-TCNQ/Ag photonic structure, for which the hybrid modes decay on time scales that are significantly faster than vibrationally mediated CT\cite{wang2021polariton,delpo2021polariton,hong2014ultrafast,chen2023recent}. Finally, we confirm that the magnitude of the anion peak scales linearly with laser fluence, ruling out many-body or nonlinear effects(Fig. S10).

These results indicate that efficient charge transfer from an organic exciton-polariton can be achieved with a significantly reduced energetic driving force as compared to the bare exciton. Molecular charge transfer is driven by the energy difference between the lowest energy donor excited state (PM605 exciton or polariton state) and the LUMO level of the acceptor (TCNQ). Outside the cavity, this value is fixed at 0.7 eV, a relatively large value compared to high performing optoelectronic systems\cite{rodovsky2013quantifying,qian_design_2018}. When strongly coupled to the photonic structure, efficient charge transfer is achieved for selective LP excitation with a driving force that is a function of the photon in-plane momentum. For example, charge transfer is greatly enhanced under PMC with 640 nm (Fig. S9) and 705 nm excitation (Fig.~\ref{fgr:MainFigCT2}b), corresponding to an energetic driving force of 0.4 eV (0.3 eV lower than the exciton energy) and 0.2 eV (0.5 eV lower), respectively.

\section{Conclusion}
In conclusion, our work underscores the critical role of cavity configuration and excitation condition in optimizing the photochemical behavior of organic exciton-polaritons. Our work confirms that polariton-enhanced charge transfer reactions are indeed achievable, but basic, long-established frameworks governing molecular systems cannot be ignored. In most systems, the key parameters are selective excitation of polariton states via energy- and momentum-matching along with hybrid-state lifetimes that significantly exceed the timescales necessary for vibrationally-mediated incoherent charge transfer. As such, Bloch surface wave systems are promising platforms to explore polariton-driven photochemistry, though other promising systems such as lattice resonances may offer similar advantages. While polariton photochemistry is not the panacea that will enable systematic explorations of all relevant excited state processes, we verify that it is achievable under specific conditions.

\bibliography{0-Reference}

\section{Materials and methods}
\subsection{\label{sec:SP}Sample preparation}
The two distributed Bragg reflectors (DBRs), composed of 6 pairs of $ \text{Si}_3\text{N}_4 (110 , \text{nm}) / \text{SiO}_2 (146 , \text{nm}) $, were prepared on a 0.15 mm thick BK7 substrate using plasma-enhanced chemical vapor deposition (PECVD), resulting in a stop-band centered at 860 nm. Subsequently, an 80 nm-thick film of PM605 on the first one and PM605-TCNQ 1:1 on the second one embedded both in Polymethyl Methacrylate (PMMA) polymer matrix were spin-coated onto the DBRs. 

The metal cavities were prepared by depositing a 2 nm Germanium layer and a 30 nm silver layer using an e-beam evaporator on a 0.15 mm thick BK7 substrate, followed by spin-coating a PM605-TCNQ blend embedded in a PMMA polymer matrix, resulting in a thickness of 150 nm.

\subsection{\label{sec: TRPPTDS}Time-resolved reflection spectroscopy}
A pump-probe reflection geometry was employed using the Kretschmann–Raether attenuated total reflection method. A Yb:KGW amplified laser system (Light Conversion, Carbide CB3) operating at a center wavelength of 1030 nm delivered pulses at a repetition rate of 90 kHz. An optical parametric amplifier (OPA) (Light Conversion, Orpheus-F) generated the pump pulse, spanning from UV to NIR. A yttrium aluminum garnet (YAG) crystal produced the probe light with a supercontinuum spectrum covering the visible (VIS) and near-infrared (NIR) ranges. In the Kretschmann-Raether configuration, both pump and probe pulses propagated through a glass prism. The pump-probe delay was mechanically controlled using a delay stage and retroreflector before generating white light. Broadband quarter-waveplates and polarizers were employed to regulate the intensity and polarization direction of both the pump and probe beams. A fast line-scan camera (Teledyne, e2V Octoplus USB) with a burst modulation scheme facilitated multichannel shot-to-shot detection.

\subsection{\label{sec:ARRS}Angle-resolved reflection spectroscopy}
The angle-resolved reflectivity was measured using a home-build Fourier space imaging setup consisting of a tungsten halogen lamp. White light from the lamp was focused onto the sample using an oil immersion objective lens (100x, 1.4 NA, Olympus). Immersion oil facilitated the attachment of the glass substrate to the objective lens. Reflected light was collected through the same microscope objective and directed to the back focal plane of the imaging objective. The light at the back focal plane was then directed through a monochromator (Acton SpectraPro SP-2500, Princeton Instruments) and detected by an electron-multiplying charge-coupled device (EMCCD) camera (PIXIS: 256, Princeton Instruments).

\subsection{\label{sec: TRPPTDS}Electrochemistry methods}
Spectroelectrochemistry measurements were conducted using a home-built setup. The sample was irradiated with collimated white light (200 nm - 2.5 $\mu$m) and the emitted light was collected with a CCD spectrometer. The sample was placed in a 2 mm electrochemistry cell equipped with a pre-assembled mesh working electrode, a Pt counter electrode, and an Ag/Ag$^+$ pseudo reference electrode. Cyclic voltammetry measurements were performed in a three-electrode configuration cell. The setup included an Ag/Ag$^+$ pseudo reference electrode (referenced to the Fc/Fc$^+$ couple), a glassy carbon working electrode, and a platinum counter electrode. Cyclic voltammetry and spectroelectrochemistry measurements were carried out in 1 mM and 0.1 mM analyte solutions, respectively, under an inert atmosphere maintained by a constant argon flux. Acetonitrile (anhydrous, 99.8\%) was used as the solvent, with [nBu$_4$N]PF$_6$ serving as the supporting electrolyte.
\end{document}


\title [Supplementary Information]{Efficient and Tunable Photochemical Charge Transfer via Long-Lived Bloch Surface Wave Polaritons}
\maketitle

 \tableofcontents
\beginsupplement

\section{\label{sec:Elevels} Energy Diagrams and Charge Transfer Pathways Outside the Cavity}
\begin{figure}[h!]
    \centering   \includegraphics[width=0.5\linewidth]{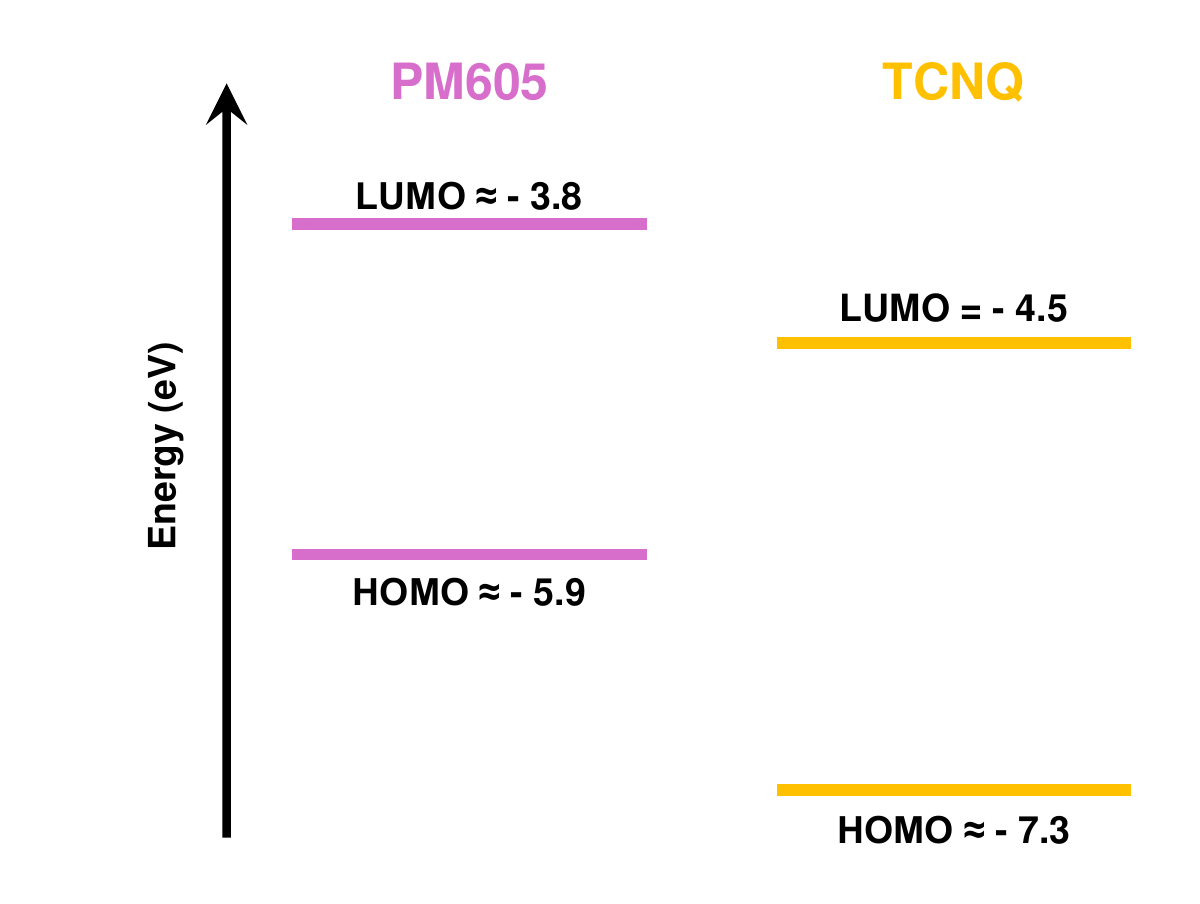} 
  \caption{\textbf{Energy-level diagram.} Energy diagram for the donor (PM605) and acceptor (TCNQ) molecules.}
  \label{fgr:HOMOLUMO}
\end{figure}

Charge transfer is initiated by the transition of electrons from the donor's highest occupied molecular orbital (HOMO) to the acceptor's lowest unoccupied molecular orbital (LUMO).
Fig.~\ref{fgr:HOMOLUMO} depicts the energy level diagram of the donor (PM605) and acceptor (TCNQ) molecules outside the cavity. As we can see, the LUMO of the donor molecules is at a higher energy level than that of the acceptor molecule, and the HOMO of the acceptor is at a significantly lower energy level. Therefore, optical excitation of the PM605-TCNQ blend will result in the transfer of charge from the donor to the acceptor via a non-radiative path.

\section{\label{sec:BarePM605}Transient Transmission of the Reference Donor Material}
\begin{figure}[h!]
    \centering   \includegraphics[width=0.75\linewidth]{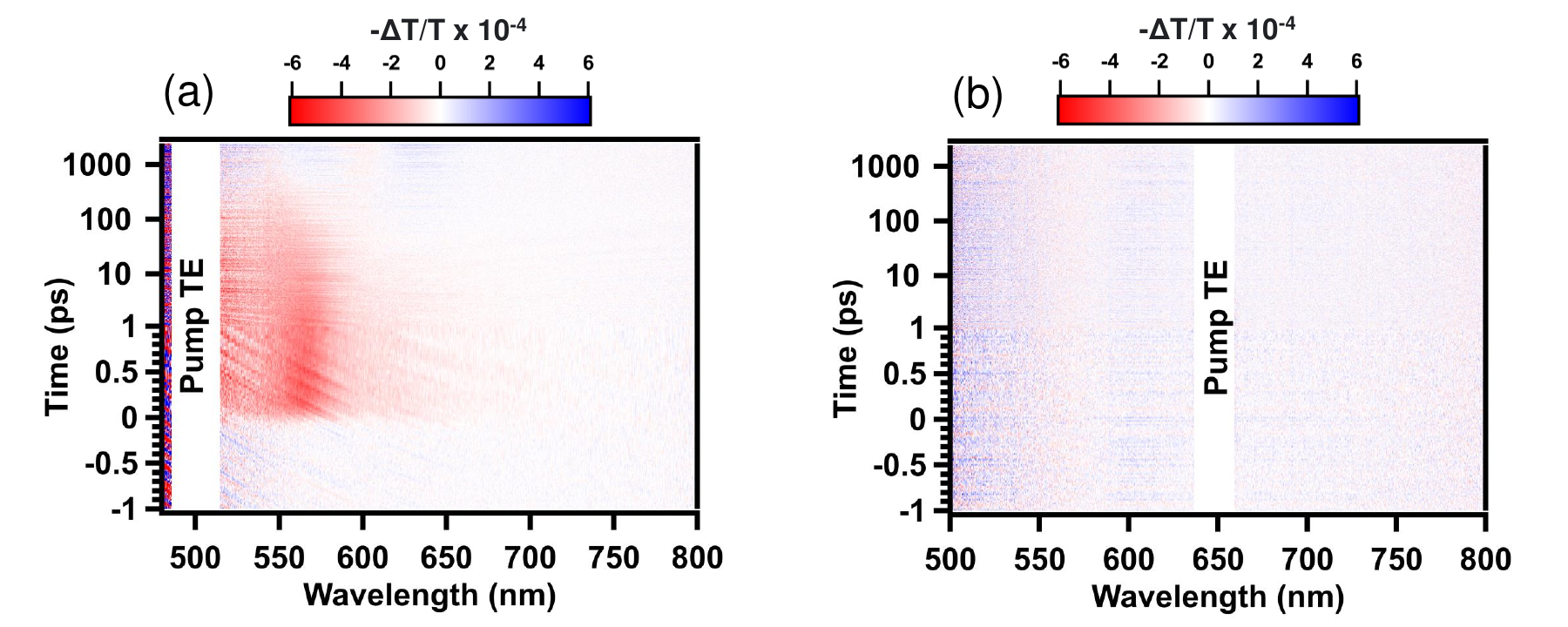} 
  \caption{\textbf{Dynamics of the bare PM605 film.} Transient transmission of the bare PM605 film following (a) 500 nm and (b) 650 nm pumping.}
  \label{fgr:Bare-DBR}
\end{figure}

The transient transmission data of the reference donor material (PM605) bare film in Fig.~\ref{fgr:Bare-DBR}a exhibits prominent negative $-\Delta \mathrm{T}/\mathrm{T}$, indicating bleaching of the ground-state absorption. However, at higher wavelengths, this effect is not observed, as the pump is no longer within the absorption range of the donor material (Fig.~\ref{fgr:Bare-DBR}b).

\section{\label{sec:PMC}Phase-Matching Condition (PMC)}
\begin{figure}[h!]
    \centering   \includegraphics[width=0.4\linewidth]{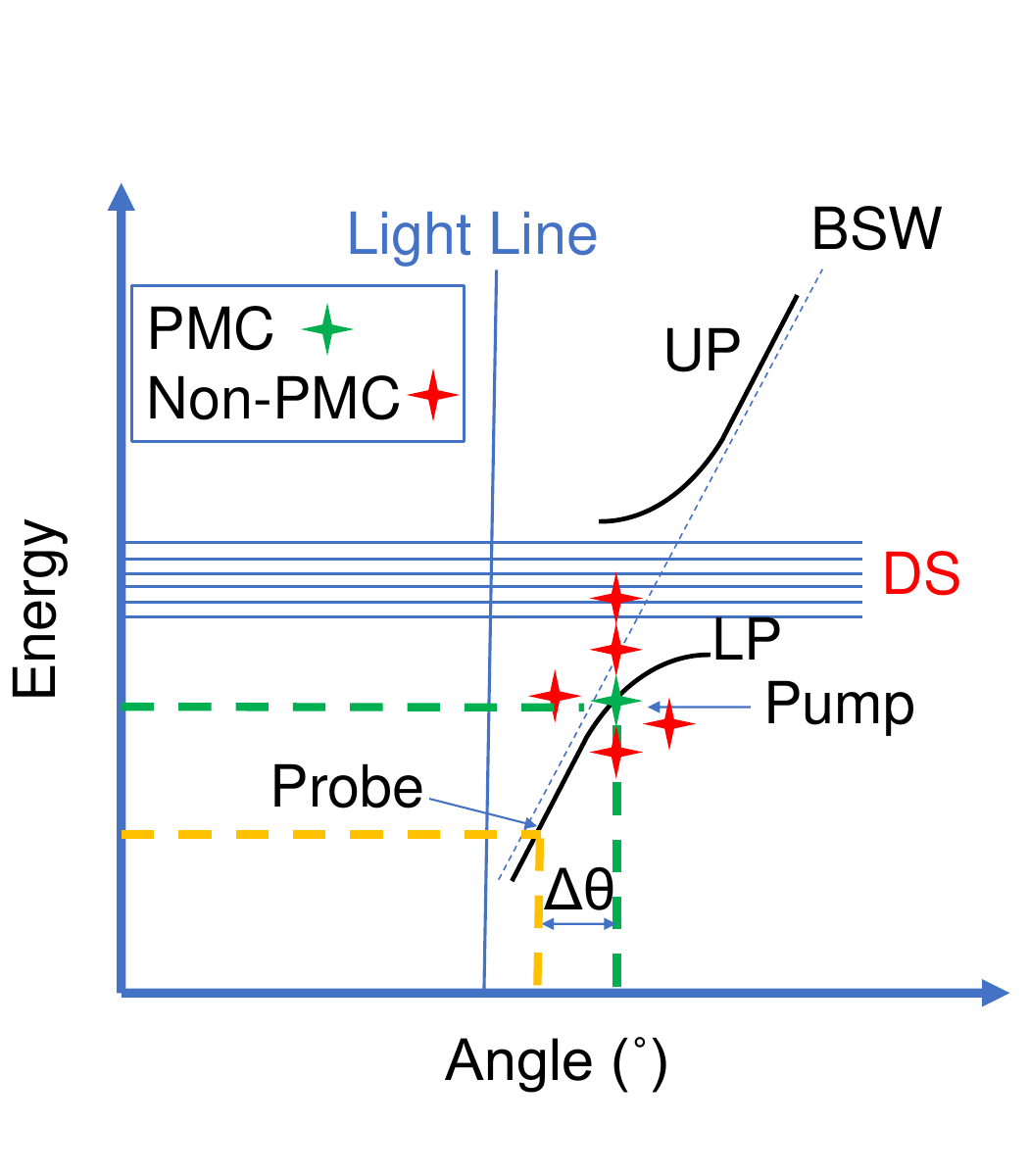} 
 \caption{\textbf{Excitation under PMC.} Schematic representation of the PMC. The condition is met when the angle (momentum) and wavelength (energy) of the pump fall within the LP branch. Otherwise, the system operates under non-phase matching conditions(Non-PMC).}
  \label{fgr:Config}
\end{figure}

For selective LP pumping under PMC, we tune both the pump's energy (wavelength) and momentum (angle) to match the LP dispersion curve, which allows us to probe the intrinsic dynamics of polaritons (Fig.~\ref{fgr:Config}). Additionally, the pump wavelength should coincide with a region of strong light emission observed in PL measurements of PM605. This approach predominantly excites LP modes while minimizing excitation of DSs.

\section{\label{sec:FlDep}Fluence-Dependent Measurements of PM605/DBR}

\begin{figure}[h!]
    \centering
    \includegraphics[width=1\linewidth]{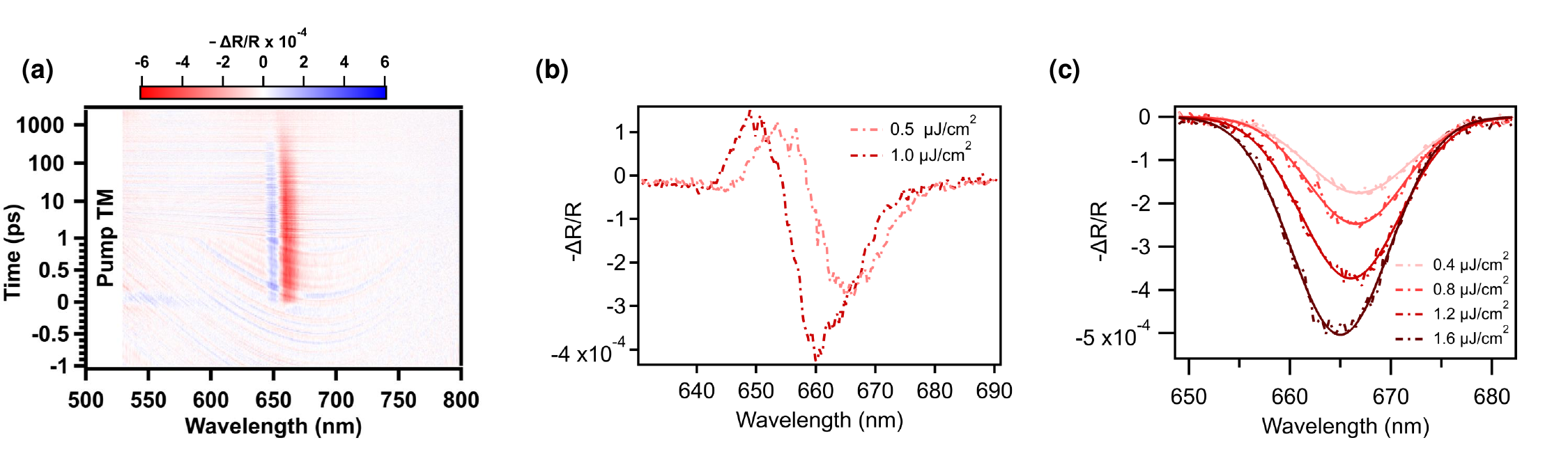} 
  \caption{\textbf{Fluence-dependent transient reflectance of PM605/DBR.}
  (a) Pumping reservoir states with TM-polarized light at 520 nm. (b) Fluence-dependent transient reflection spectra obtained by exciting reservoir states at 520 nm and (c) polariton states under PMC. In both cases, as the fluence increases, a distinct blue shift in the signal is observed, a characteristic feature of strongly coupled systems.}
  \label{fgr:SI-powerdepLonglived}
\end{figure}

To study the dynamics of DSs in PM605/DBR configuation, we employ a pump beam with TM polarization. Since the DBR reflects TE-polarized light but not TM polarization at wavelengths resonant with the exciton material transition, the TM polarization can be absorbed by the excitonic material. Consequently, the ground state bleaching (GSB) accompanied by a blue shift of the excitonic material can be observed in the probe beam. This observation can be detected as a combination of a derivative shape and a negative signal in the transient reflection (Fig.~\ref{fgr:SI-powerdepLonglived}a). In this experiment the system is solely probed using TE polarization, given that the polariton is generated through the TE mode.

The power dependence measurements of DS pumping shows a blue shift in the transient signal and an increase in the GSB as we increase the power. This shift is attributed to the decrease in Rabi splitting, as Rabi splitting is given by \(\Omega \propto \sqrt{N}\), where the photobleaching of excitons increases with increasing pump power (Fig.~\ref{fgr:SI-powerdepLonglived}b). A similar effect, as shown in Fig.~\ref{fgr:SI-powerdepLonglived}c, can be observed for the long-lived signal under the PMC.  Furthermore, the fluence-dependent photoluminescence measurements demonstrate an approximately linear increase in the PL data with increasing pump power (Fig.~\ref{fgr:SI-PL-powerdep}), where no nonlinear or threshold-like behavior is observed. This implies that there are no signs of lasing or nonlinear behavior.

\begin{figure} [h!]
    \centering
    \includegraphics[width=0.8\linewidth]{Fig S5.pdf}    
  \caption{\textbf{Fluence-dependent photoluminescence of PM605/DBR.} (a) achieved by exciting polariton states with TE-polarized light under PMC.(b) The integrated emission intensity as a function of fluence.
}
  \label{fgr:SI-PL-powerdep}
\end{figure}

\section{\label{sec:DiffAng} PMC for Different Angles Between the Pump and the Probe Beams}
\begin{figure}
    \centering    \includegraphics[width=1\linewidth]{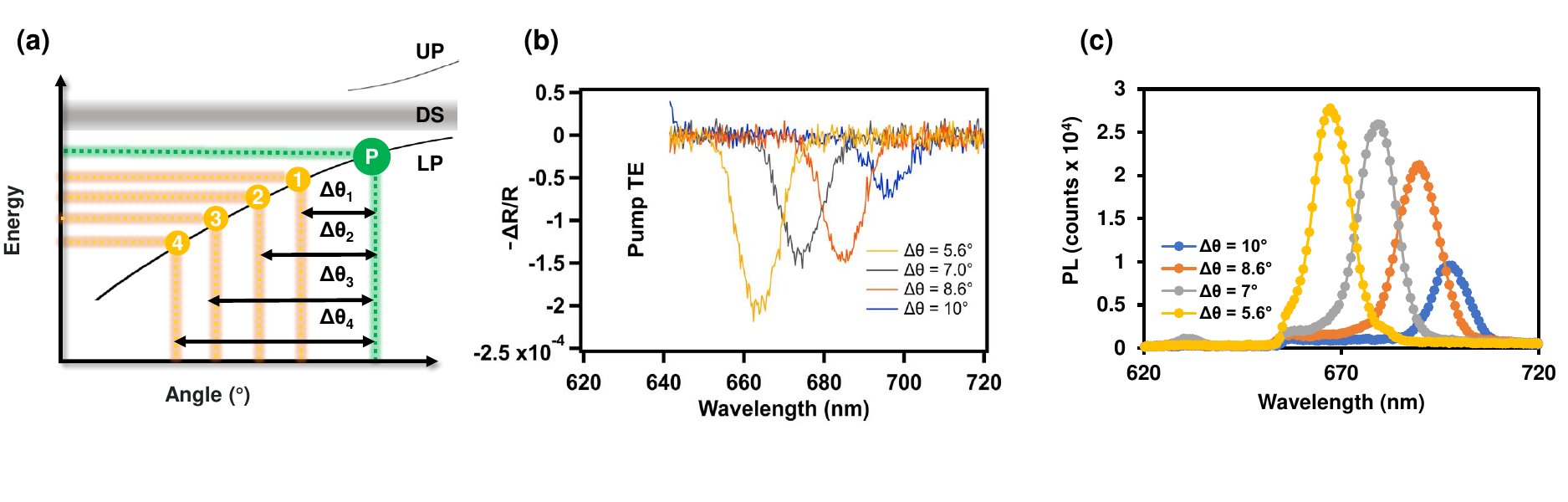}  
  \caption{\textbf{PMC for different angles between the pump and the probe.} (a) Schematic illustrating the PMC for four different angles between the pump and the probe beams. (b) Transient reflectance spectra of the long-lived signal and (c) steady-state emission for the four angles.}
  \label{fgr:difftheta}
\end{figure}

To illustrate the origin of this long-lived signal, we compare the steady-state emission and transient reflection signals. For this purpose, we pump the polariton at a fixed wavelength (633 nm) and probe at four different angles: $d\theta = 5.6^\circ$, $7^\circ$, $8.6^\circ$, and $10^\circ$ (Fig.~\ref{fgr:difftheta}). As we increase the angle between the pump and the probe, the long-lived transient signal generally ($-\Delta R/R$) decreases (Fig.~\ref{fgr:difftheta}b), which is consistent with the decrease in the steady-state emission signal (Fig.~\ref{fgr:difftheta}c). The similarity between the LP transient reflection data and the lower polariton PL data indicates that the relaxation of the  state mainly occurs through the emission of photons. This implies that other relaxation pathways, such as non-radiative decay, play a less significant role.

Additionally, the simultaneous increase in quality factor and decrease in mode volume of BSW enhance the Purcell effect, resulting in increased spontaneous emission. The amplified spontaneous emission directly impacts stimulated emission, potentially leading to stimulated emission without a threshold. Moreover, this illustrates the spectrum of the bsw-polariton, functioning as a filter that enables the excitation and probing of both excitons and polaritons. 

\section{\label{sec:PW} Charge Transfer kinetics}
\begin{figure}[h!]
    \centering
    \includegraphics[width=0.8\linewidth]{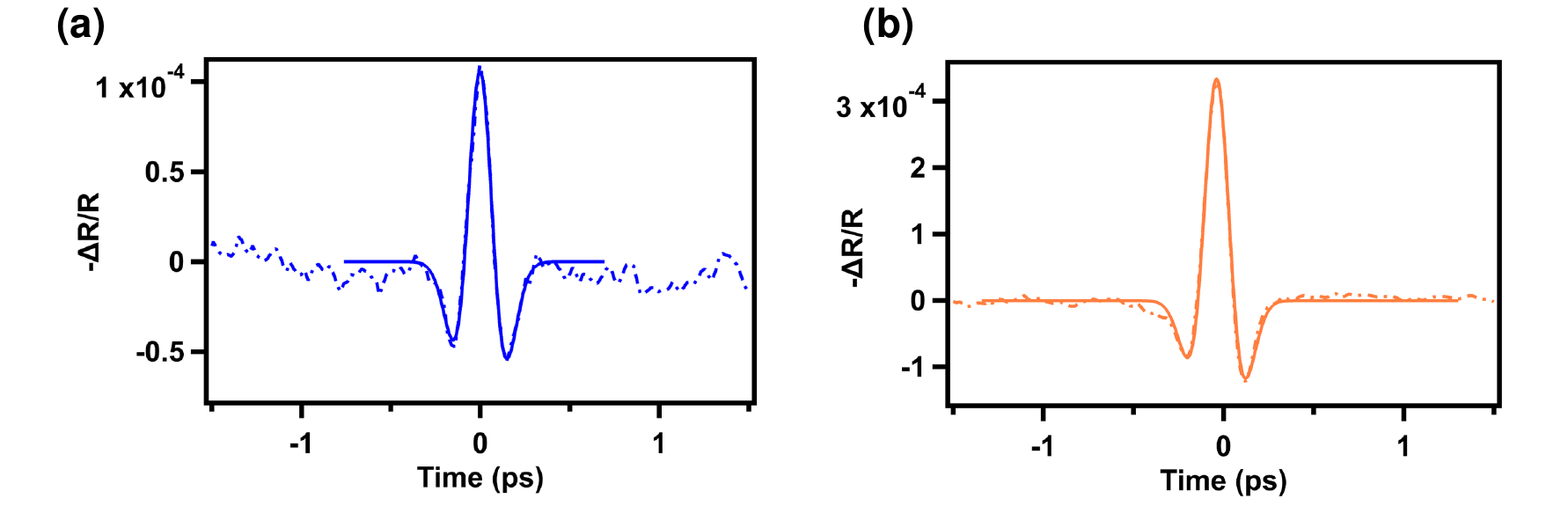}    \caption{\textbf{Pulse width calculation.} Fitting cross-phase modulation data at early times for excitation wavelengths (a) 510 nm and (b) 710 nm, respectively. In both cases, the pulse width is 0.09 ps.}
  \label{fgr:PulseWidth}
\end{figure}

To calculate the pulse width and determine an accurate time zero, we utilize the artifact at the the early times and fit it to three Gaussian functions. This fitting combines Gaussian functions and Hermite-Gaussian functions to accurately represent the temporal profile of the pulses. Fig.~\ref{fgr:PulseWidth} illustrates these two fittings for excitation at (a) 510 nm and (b) 710 nm, respectively.

\begin{figure}[h!]
    \centering
    \includegraphics[width=0.7\linewidth]{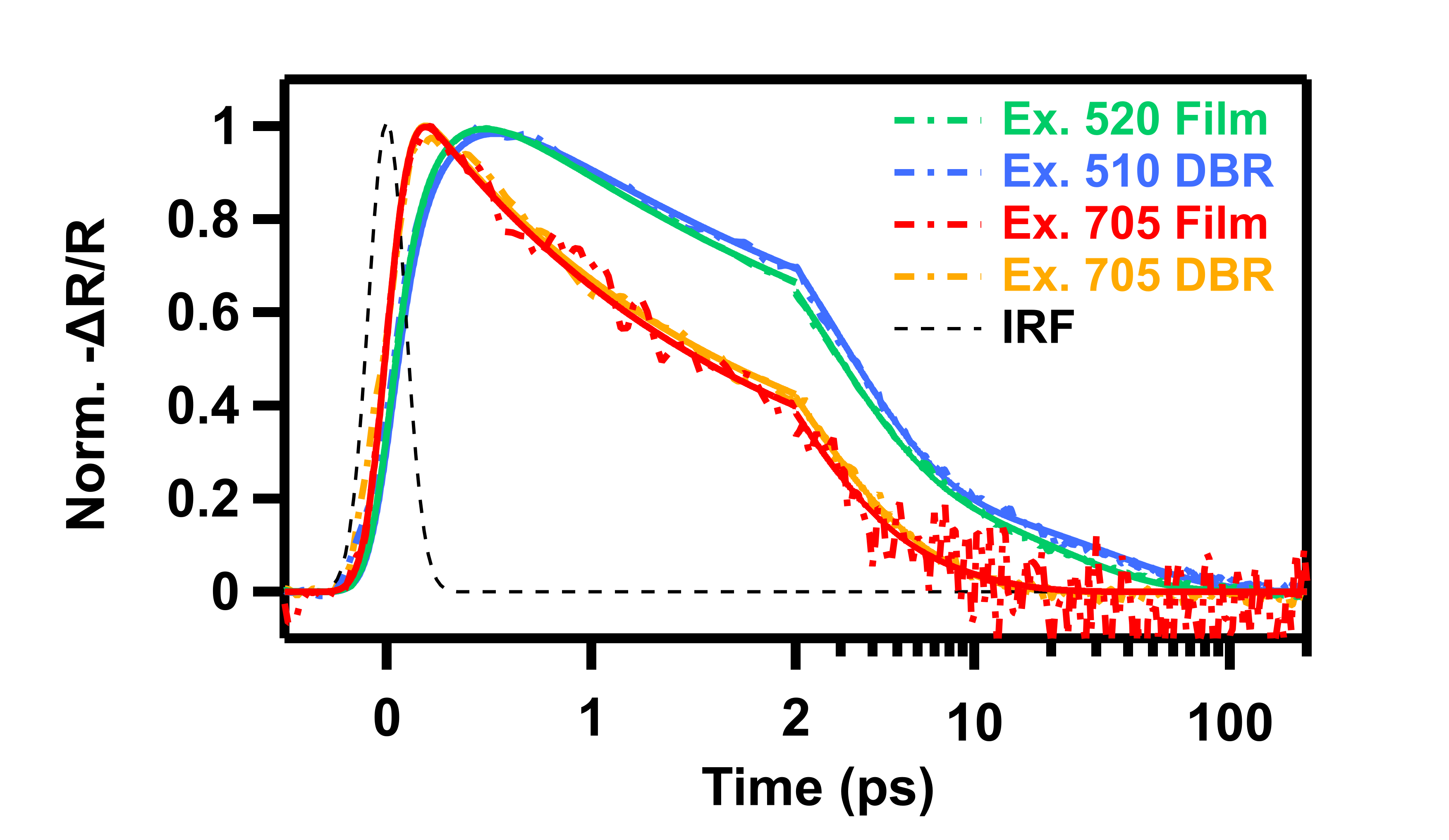}    \caption{\textbf{CT kinetics within PM605-TCNQ/DBR.}
Kinetic data obtained from exciting the PM605-TCNQ film at 520 nm and 705 nm, as well as the PM605-TCNQ/DBR at 510 nm and 705 nm, are depicted alongside instrument response functions (IRF) and multi-exponential fittings (solid lines). While excitation at lower wavelengths exhibits a gradual rise and decay, which is slower than the IRF, excitation at higher wavelengths results in faster anion formation and decay.}
  \label{fgr:CTSign}
\end{figure}

The kinetic traces in Fig.~\ref{fgr:CTSign} exhibit a gradual increase under lower wavelength excitation (510nm film and 520 nm DBR), exceeding the instrument response function (IRF). This gradual rise near the anion peak provides compelling evidence suggesting the formation of CT states. In contrast, under higher wavelength excitation, the rise occurs faster for both DBR and bare film. The consistency in rise time between DBR and film suggests that the rise is not merely a broadened pulse width due to glass dispersion. Furthermore, it indicates that the bsw polariton does not hinder CT, possessing the same lifetime and kinetics as the bare dye, albeit enhanced and controllable for polariton. Additionally, relaxation is prolonged for DS excitation, indicating a slight deviation in the decay pathway compared to LP excitation due to differing driving forces.

\section{\label{sec:AgvsDBR} Comparing Metal and DBR under Identical Excitation}
\begin{figure}[h!]
    \centering
    \includegraphics[width=0.5\linewidth]{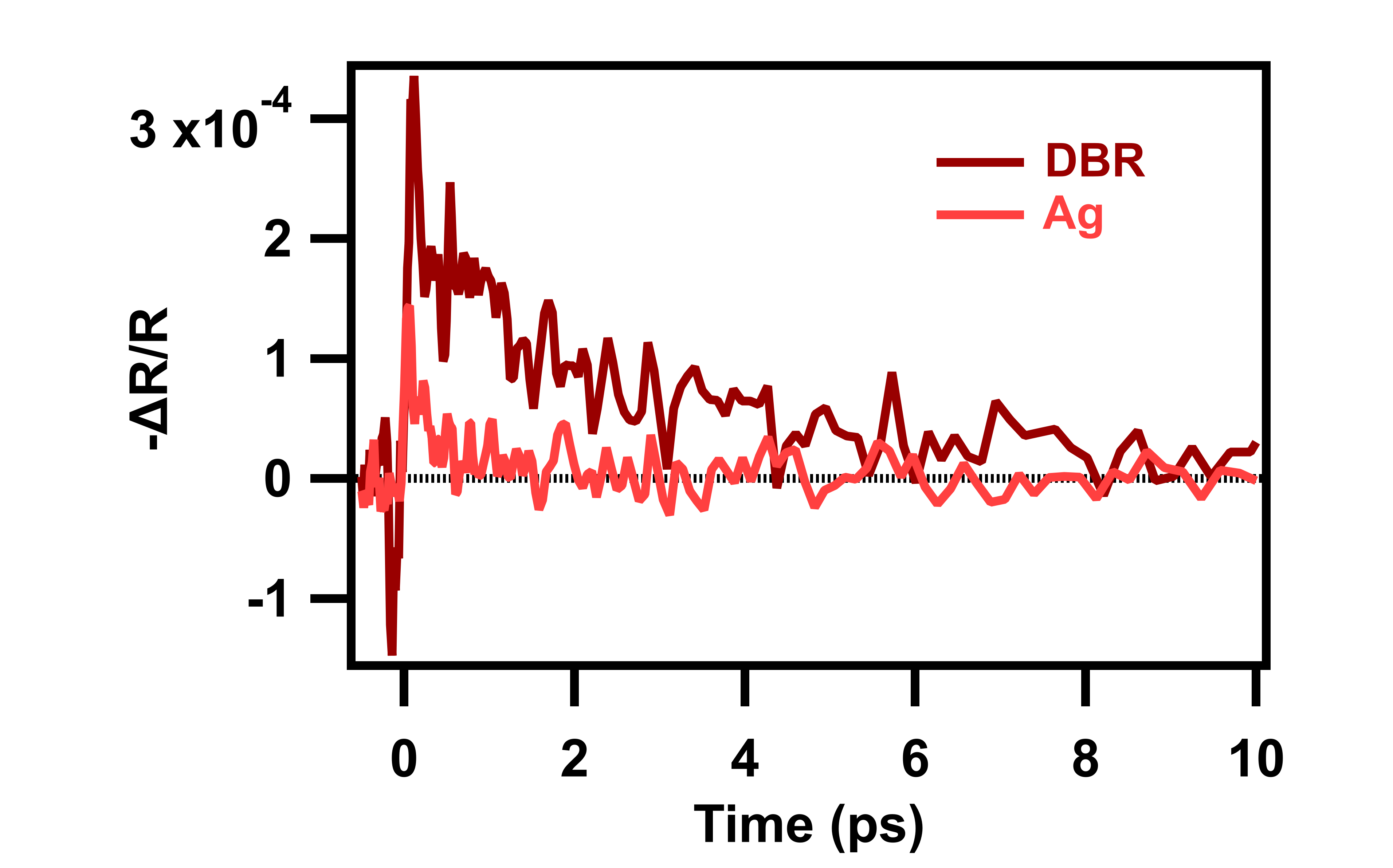}    
  \caption{\textbf{Metal vs. DBR cavity.} Comparing the Kinetics of PM605-TCNQ/Ag and PM605-TCNQ/DBR under the same excitation condition. PM605-TCNQ/Ag exhibits no discernible CT, while PM605-TCNQ/DBR shows a pronounced CT signal.
}
  \label{fgr:WGvsBSW}
\end{figure}

Fig.~\ref{fgr:WGvsBSW} compares the transient reflection data of the LP mode in PM605-TCNQ/Ag with the PM605-TCNQ/DBR configuration. As you can see, there is no detectable signal similar to the charge transfers in the Ag cavity, while in the DBR, we can observe a signal with kinetics similar to that of bare PM605-TCNQ. This indicates charge transfer in higher lifetime cavities (DBR) and no charge transfer in lower lifetime and dissipative cavities (metals).

\section{\label{sec:FlDep}Fluence-Dependent Measurements of PM605-TCNQ/DBR}
\begin{figure}[h!]
    \centering
    \includegraphics[width=0.5\linewidth]{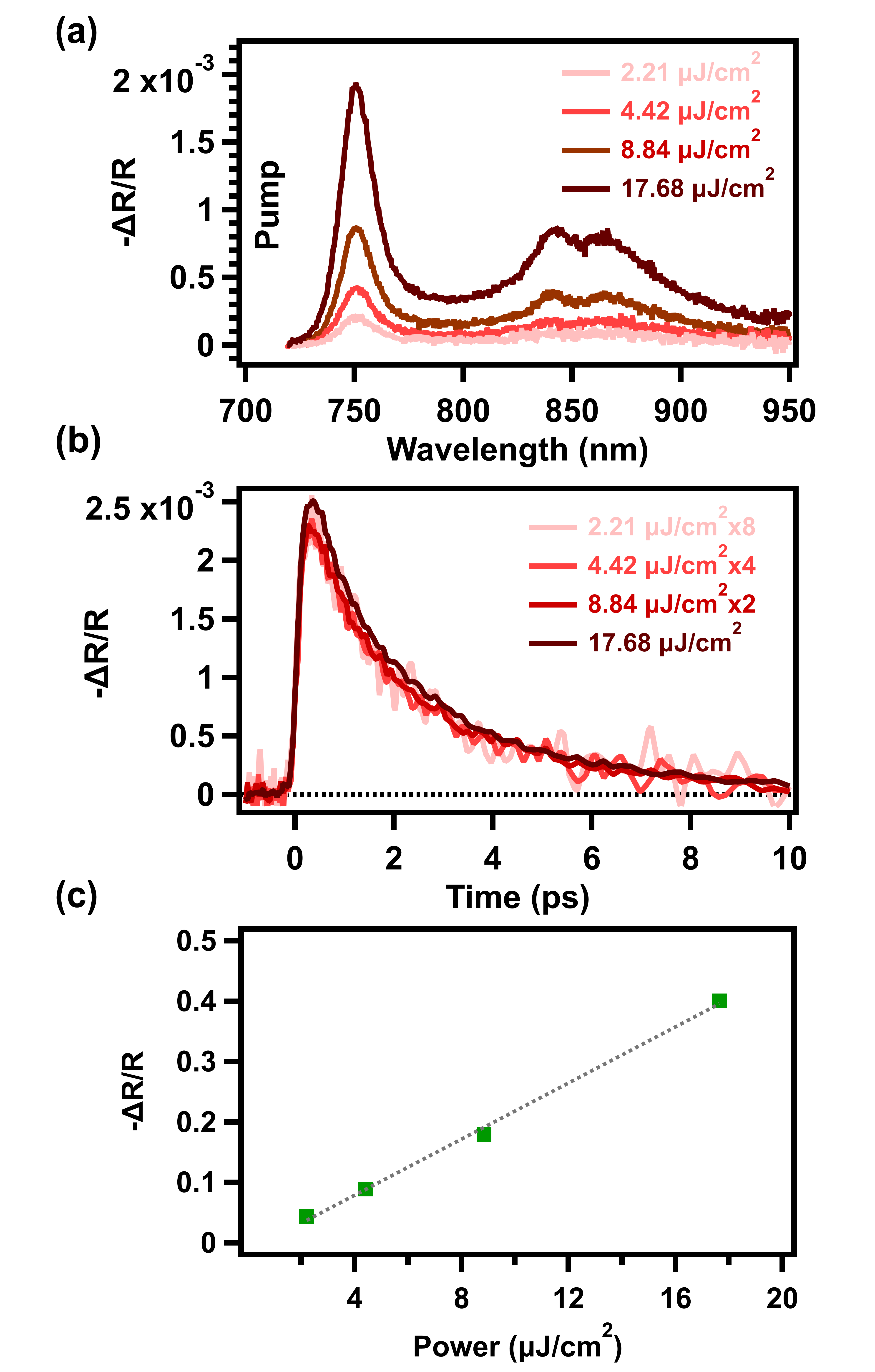}    
  \caption{\textbf{Fluence-dependent transient reflectance of PM605-TCNQ/DBR.} (a) Spectra and (b) kinetics of PM605-TCNQ/DBR obtained by exciting LP states at 700 nm. (c) The integrated $-\Delta R/R$ as a function of fluence under PMC.
}
  \label{fgr:SI-blend-powerdep}
\end{figure}

We have performed the Fluence-Dependent Measurements of PM605-TCNQ/DBR and the spectra and  kinetics of PM605-TCNQ/DBR obtained by exciting LP states at 705 nm are depicted in Fig.~\ref{fgr:SI-blend-powerdep}.
The transient reflection data of the blend (PM605-TCNQ/DBR) does not exhibit a nonlinear increase with fluence, as shown in the transient spectrum and kinetics (Fig.~\ref{fgr:SI-blend-powerdep}). Therefore, we cannot assign this process to Auger recombination or nonlinear effects.

